\documentclass[10pt,conference]{IEEEtran}
\usepackage{cite}
\usepackage{amsmath,amssymb,amsfonts}
\usepackage{algorithmic}
\usepackage{graphicx}
\usepackage{textcomp}
\usepackage{xcolor}
\usepackage[hyphens]{url}
\usepackage{fancyhdr}
\usepackage{hyperref}
\usepackage{siunitx}
\usepackage{booktabs}
\usepackage{multicol}
\usepackage{multirow}
\usepackage{arydshln}
\usepackage{subcaption}
\usepackage{verbatim}

\usepackage{tikz}

\usepackage{pifont}
\newcommand{\cnum}[1]{%
\tikz[baseline=(char.base)]{
\node[draw,circle,fill=black,text=white,
inner sep=0.6pt,font=\scriptsize\bfseries] (char) {#1};}}

\usepackage{ifthen}
\usepackage[dvipsnames]{xcolor}
\usepackage{mathabx}
\newboolean{showcomments}
\setboolean{showcomments}{false}

\makeatletter
\newcommand{\mynote}[3]{%
  \ifthenelse{\boolean{showcomments}}{%
   \fbox{\bfseries\sffamily\scriptsize#1}%
   {\small$\blacktriangleright$\textsf{\emph{\color{#3}{#2}}}$\blacktriangleleft$}}%
  {%
   \@bsphack
   \@esphack
  }%
}
\makeatother

\newcommand{\jp}[1]{\mynote{JP}{#1}{gray}}
\definecolor{asparagus}{rgb}{0.53, 0.66, 0.42}

\makeatletter
\long\gdef\rev#1{#1}                  
\makeatother
\DeclareSIUnit{\ops}{ops}

\newcommand{\paperName}{BROAD}
\newcommand{\paperLongName}{Bit-Parallel Row-Oriented Arithmetic in DRAM}
\title{\paperName: \paperLongName}

\author{%
\IEEEauthorblockN{%
William C. Tegge\IEEEauthorrefmark{1},
Jo\~{a}o Paulo Cardoso de Lima\IEEEauthorrefmark{2},
Shouzhi Fang\IEEEauthorrefmark{1},
Jeronimo Castrillon\IEEEauthorrefmark{3},
Alex K. Jones\IEEEauthorrefmark{1}}
\IEEEauthorblockA{\IEEEauthorrefmark{1}Syracuse University, Syracuse, NY, USA\\
\{wtegge, sfang18, akj\}@syr.edu}
\IEEEauthorblockA{\IEEEauthorrefmark{2}Friedrich-Alexander-Universit\"{a}t Erlangen-N\"{u}rnberg, Erlangen, Germany\\
joao.lima@fau.de}
\IEEEauthorblockA{\IEEEauthorrefmark{3}TU Dresden, Dresden, Germany\\
jeronimo.castrillon@tu-dresden.de}
}

\begin{document}

\maketitle

\begin{abstract}
Processing-using-memory (PUM) architectures perform computation directly within DRAM to reduce costly data movement between memory and processors. Because charge-sharing operations are confined to individual bitlines, existing charge-sharing-based DRAM-PUM architectures reorganize data into column-oriented, bit-serial representations. This organization is fundamentally incompatible with the row-oriented, word-parallel layouts used by conventional processors and accelerators, requiring expensive data-layout transformations whenever computation transitions between PUM and conventional execution.
In this paper, we present \paperName{}, or \paperLongName{}. \paperName{} augments the DRAM subarray with two lightweight extensions: migration cells that enable localized horizontal data movement between neighboring bitlines and inversion cells that provide efficient in-array logical inversion. These primitives enable \paperName{} to operate directly on row-parallel, bit-parallel data, preserving CPU-compatible layouts while exploiting the massive parallelism of the DRAM subarray. In particular, \paperName{} demonstrates that localized horizontal communication is sufficient to realize shallow arithmetic networks and efficient parallel reduction for multiplication, reducing arithmetic latency while preserving throughput and eliminating costly data-layout transformations, all while maintaining the conventional DRAM array organization.
We demonstrate the feasibility and overhead of augmenting DRAM subarrays with migration and inversion cells through detailed transistor-level layout and SPICE-validated circuit simulations. Using the \paperName{} compiler it is possible to evaluate the performance and data reorganization tradeoffs to ensure the best execution across combined CPU and PUM. Evaluating \paperName{} on 19 MLPerf benchmarks, there is a 5.9$\times$ higher end-to-end performance compared to SIMDRAM for DDR4 PUM execution.
\end{abstract}


\section{Introduction}
\label{sec:intro}

Modern computing systems rely on main memory to orchestrate data movement between processors and increasingly diverse compute accelerators. Processing-using-memory (PUM) has emerged as a promising approach to overcome the memory bandwidth bottleneck by performing computation directly within DRAM, eliminating much of the costly data movement between memory and compute units~\cite{seshadri2017ambit,yuksel2024functionally,hajinazar2021simdram,oliveira2024mimdram,elp2im,de2024count2multiply,li2017drisa,manyrowact}. However, most DRAM-PUM architectures reorganize data into column-oriented, bit-serial representations to exploit the massive bit-level parallelism of the memory array. While effective for in-memory Boolean computation, these organizations differ fundamentally from the row-parallel, word-oriented layout used by conventional processors, complicating CPU-PUM interoperability and often requiring expensive data-layout transformations.

The motivation for column-oriented organization stems from the need for fine-grained access to individual operand bits. Arithmetic operators are built from Boolean primitives and, more recently, majority logic~\cite{soeken2017exact,hajinazar2021simdram,oliveira2024mimdram}, repeatedly combining bits across significance positions to construct carry chains, partial products, and prefix networks. Commodity DRAM isolates neighboring columns while sharing bitlines, so no native mechanism exists for communication between adjacent bit positions. Existing DRAM-PUM therefore places each bit position in a separate row~\cite{seshadri2017ambit,elp2im,yuksel2024functionally,li2017drisa,hajinazar2021simdram,de2024count2multiply,oliveira2024mimdram} and realizes arithmetic through logical shifting, selecting different rows rather than physically moving data. This buys complete flexibility to align arbitrary bit positions, but carry propagation, partial-product accumulation, and prefix computation all operate over structured local neighborhoods. Restricted horizontal movement may therefore suffice, and the flexibility of unrestricted logical shifting may not justify a layout incompatible with the row-major organization processors use.

Prior work~\cite{akin2015data,li2017drisa,CORUSCANT} has recognized the importance of efficient horizontal data movement in DRAM-PUM, for applications including genomics~\cite{Sieve}, cryptography~\cite{fang26-aes} and hyperdimensional inference~\cite{HDC-Brain}. It is also important to distinguish the \emph{data layout} (how operand bits are arranged in the array) from the \emph{execution style} (whether carries propagate serially or in parallel). Proteus~\cite{de2025proteus} shows that bit-parallel algorithms such as carry-lookahead addition can execute on vertically organized data by distributing operand bits across subarrays and exploiting subarray-level parallelism, so algorithmic parallelism is largely decoupled from layout. The physical organization nonetheless remains bit-sliced, still requiring transposition at the row-major host boundary. A complementary approach is to run bit-parallel algorithms directly on a row-major layout, needing only local horizontal movement for carry propagation while preserving CPU-compatible data organization and eliminating transposition entirely.

In this paper we present \paperName{}, or \paperLongName{}, a DRAM-PUM architecture that exploits the latency advantage of bit-parallel computation while retaining enough parallelism to match bit-serial throughput. It augments the subarray with two lightweight primitives. \textit{Migration cells}~\cite{lu2015improving}, placed in pairs at distinct offsets, move data horizontally between neighboring bitlines; \textit{inversion cells}, a migration-cell-derived structure at the subarray extremities, give controlled access to the complementary side of the sense amplifier so inverted values can be restored to their original bitlines. Both preserve conventional array organization and density.

On these primitives \paperName{} builds Kogge--Stone prefix addition in $O(\log n)$ \rev{logic} depth, mapping each prefix level to power-of-two shifts followed by majority-based \texttt{AND}/\texttt{OR}. \rev{Because \texttt{RowShift} moves one column per operation, the shift cost is linear in $n$ rather than logarithmic (\autoref{sec:limitations}).} Multiplication generates partial products by migration-cell broadcast, compresses them with carry-save reduction using only single-bit horizontal shifts, and resolves the result with one prefix addition---all on row-parallel data, preserving CPU-compatible layouts throughout. Because both layouts are natively executable, the \paperName{} compiler performs a global optimization over the dependence graph, jointly selecting data layouts and arithmetic algorithms to minimize the combined cost of computation and layout conversion.

In particular, the contributions of \paperName{} are as follows:
\begin{enumerate}



\item We introduce a fundamentally new migration-cell-based computational capability that enables parallel pairwise bit interaction between adjacent bit positions within a row. This allows adjacent bit pairs to simultaneously participate in majority and clone operations with one or more aligned rows, which cannot be realized by augmenting the local row buffer with shifting circuitry~\cite{li2017drisa}.


\item We develop a shallow row-parallel arithmetic framework that preserves conventional processor-compatible data layouts. The framework exploits parallel pairwise bit interaction to realize partial-product masking, \texttt{MAJ3}/\texttt{MAJ5}-based carry-save reduction, and logarithmic \rev{logic-}depth prefix addition.


\item We develop the first compiler for heterogeneous bit-serial/bit-parallel DRAM-PUM that globally co-optimizes arithmetic algorithm selection and data layout, minimizing the combined cost of PUM execution and host-side data reorganization across the dependence graph. \rev{On kernels whose arithmetic operations prefer different layouts, this joint selection lowers the latency by $1.8\times$ over the fixed layout compilation.}

\item We evaluate the tradeoffs between row-parallel and bit-serial DRAM-PUM organizations, demonstrating that row-parallel execution preserves CPU and accelerator interoperability, eliminates costly data-layout transformations, substantially reduces arithmetic latency, and maintains throughput competitive with state-of-the-art bit-serial DRAM-PUM architectures.
\end{enumerate}
\paperName{} provides a 5.9$\times$ performance improvement over SIMDRAM~\cite{hajinazar2021simdram} for host-PUM edge-class cooperative execution of MLPerf workloads.

\section{Background and Motivation}

\subsection{DRAM Organization}
\label{sec:bg-dram}

\begin{figure}
    \centering
    \includegraphics[width=\linewidth,trim={1cm 5cm 1cm 9.5cm}, clip]{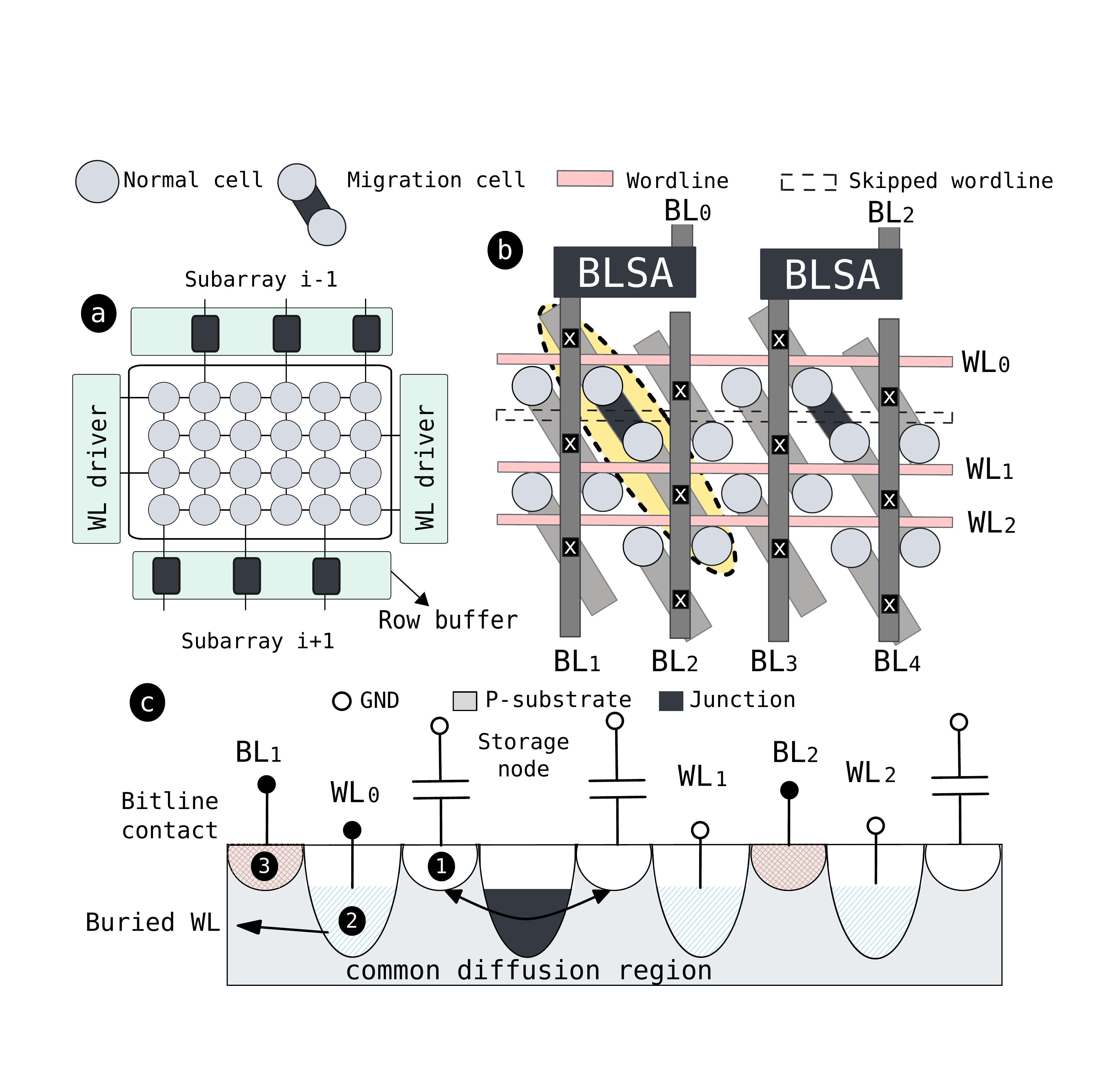}
    \vspace{-2em}
    \caption{(a) DRAM open-bitline architecture, (b) migration cell design, (c) cross-section of a saddle-fin transistor.}
    \label{fig:dram_orga}
\end{figure}

A DRAM bank is divided into subarrays composed of MATs (\autoref{fig:dram_orga}a). Each cell holds one bit as charge on a capacitor gated by a wordline (WL), and the cells on a bitline (BL) share a BLSA at the subarray boundary that resolves an activated cell's charge against the BL's precharged $\frac{1}{2}V_{DD}$ level~\cite{seshadri2017ambit,kim2012case}. High-density parts use an open-bitline 6$F^2$ organization~\cite{DRAMScope,hong2023dsac} in which each BLSA is shared between adjacent subarrays, pairing BL $i$ of one with BL $i{+}1$ of its neighbor~\cite{jacob2010memory}. \autoref{fig:dram_orga}c shows the cross-section, with a saddle-fin transistor~\cite{park2006highly} driving a storage node \ding{182} under a buried wordline~\cite{schloesser20086f} \ding{183} and two cells sharing a BL contact \ding{184}.

\subsection{Processing-Using-Memory (PUM)}
\label{sec:bg-pim}

PUM computes inside the array by repurposing the charge-sharing behavior of DRAM cells and BLSAs. RowClone~\cite{seshadri2018rowclone} issues back-to-back \texttt{ACT} commands to a source and a destination row without an intervening \texttt{PRE}, driving the BLSA's latched value into the destination cells and copying an entire row in a single DRAM timing cycle.

Ambit~\cite{seshadri2017ambit} builds on this to realize Boolean logic in-subarray using two out-of-specification sequences: \texttt{AP} (Activate--Precharge with multi-row activation) and \texttt{AAP} (Activate--Activate--Precharge). It exposes three primitives: \texttt{RowClone}, bitwise \texttt{MAJ3} via three-row activation, and bitwise \texttt{NOT} via dual-contact cells on the complementary bitline. SIMDRAM~\cite{hajinazar2021simdram} maps complex operations to optimized \texttt{MAJ3}/\texttt{NOT} sequences using majority-inverter graph synthesis~\cite{MAJ-SYN}. Activating more than three rows is also feasible in commodity DRAM~\cite{manyrowact,Mustafa20MAJ5Adder}, enabling \texttt{MAJ5}, which realizes \texttt{XOR} efficiently as $a \oplus b = \texttt{MAJ5}(a, b, 0, \bar{t}, \bar{t})$ where $t = \texttt{MAJ3}(a, b, 0)$. The full adder follows the same pattern with $t$ as the carry: $C_i = t = \texttt{MAJ3}(a_i, b_i, c_i)$ and $S_i = a_i \oplus b_i \oplus c_i$. These primitives form the instruction set for all PUM computation in this work.

The feasibility of bridging adjacent saddle-fin transistors to implement a dual-ported \emph{migration cell} was previously demonstrated in~\cite{lu2015improving}. As illustrated in~\autoref{fig:dram_orga}b, the migration cell replaces a conventional DRAM cell at periodic locations along each bitline. Each migration cell connects to two neighboring bitlines through separate access transistors ($WL_0$ and $WL_1$) sharing a common storage node, allowing either bitline to access the same temporary storage. This shared node forms the basis for the \texttt{RowShift}, \texttt{RowBroadcast}, and \texttt{RowInvert} mechanisms described in~\autoref{sec:extensions}. \rev{The migration cell bridges two tilted active-area islands, an arrangement that modern DRAM already uses. Angled active areas are what allow the 6$F^2$ cell to reach its density target, and they have been standard in volume production for more than fifteen years~\cite{park2006highly,schloesser20086f}. Samsung's honeycomb structure at 20\,nm is one production instance~\cite{Park201520nmDA}, and die teardowns confirm the arrangement across vendors and process nodes~\cite{james2010recent}. The migration cell therefore adds a connection between structures the process already builds, rather than introducing a new device. A standard CMOS logic process offers a second path as embedded DRAM, which is what the layout evaluation in~\autoref{sec:experimental-setup} demonstrates.}

\subsection{Bit-Serial and Bit-Parallel Algorithms}
\label{sec:ks-background}

\noindent\textbf{Bit-Serial (BS) Addition.} Ripple-carry addition propagates carries sequentially across $n$ bit positions, requiring $O(n)$ depth regardless of data organization~\cite{parhami2010computer}. In a bitwise-SIMD PUM setting this dependency is mapped \emph{across} data elements: one full-adder step executes simultaneously over an entire row, each bit position resolved through a fixed sequence of $c$ primitive operations that depends on the primitive decomposition and target. This trades per-element latency for aggregate throughput~\cite{farzaneh2024sherlock,leitersdorf2023aritpim}.

\noindent\textbf{Bit-Parallel (BP) Addition.}
 Parallel prefix addition reduces this to $O(\log n)$ by expressing carry computation as an associative operator over \emph{generate} ($G$) and \emph{propagate} ($P$) signal pairs~\cite{parhami2010computer}. At each position $i$: $G_i^{(0)} = a_i \cdot b_i$ and $P_i^{(0)} = a_i \oplus b_i$.
The prefix operator combines these signals over progressively wider neighborhoods, $(G, P)_{i:j} = (G_{i:k} + P_{i:k} \cdot G_{k-1:j},\; P_{i:k} \cdot P_{k-1:j})$.
The Kogge--Stone (KS) construction~\cite{koggestone} evaluates this prefix in $\lceil \log_2 n \rceil$ levels, where level $\ell$ computes $G_i^{(\ell)} = G_i^{(\ell-1)} + P_i^{(\ell-1)} \cdot G_{i-2^{\ell-1}}^{(\ell-1)}$ and $P_i^{(\ell)} = P_i^{(\ell-1)} \cdot P_{i-2^{\ell-1}}^{(\ell-1)}$.
Each level requires only shifts at power-of-two distances and \texttt{AND}/\texttt{OR} operations; a structured, local communication pattern. In \autoref{sec:rapid-addition} we show how \paperName{} maps this construction to migration-cell shifts and majority-based logic.

\subsection{Limitations of Existing PUM Architectures}

Charge-sharing PUM computes only vertically along bitlines, with no mechanism to move data horizontally between neighboring columns. Existing architectures therefore adopt a vertical layout, placing each operand bit in a separate row so arithmetic proceeds by row-index remapping. This exposes massive column-level parallelism but is incompatible with the row-parallel layout CPUs and accelerators use, forcing transposition at every PUM boundary.

\rev{Prior work recognized the need for horizontal data movement to escape this restriction and proposed dedicated shifting logic co-located with the BLSAs, enabling bit-parallel computation directly on row-oriented data~\cite{li2017drisa}. Unfortunately, this places logic devices in the DRAM array periphery, where only a DRAM process is available. DRISA's add-on logic circuits consequently occupy 24\% of die area in its NOR-based variant and 51\% in its adder-based variant~\cite{li2017drisa}.}
Hi-Fi~\cite{marazzi2024hifi} demonstrated 
model-based estimates can be off by up to 9$\times$ in the worst case, suggesting the true cost of dedicated shift logic may actually be even higher.

By bridging the saddle-fin transistor islands at the subarray extremities~\cite{lu2015improving}, 
\paperName{} proposes 
repurposing the existing DRAM cell structure at less than 1\% area overhead~\cite{tegge2026low} rather than introducing bulk logic transistors, to provide the horizontal communication bit-parallel arithmetic requires.

\section{Lightweight Hardware Extensions for PUM}
\label{sec:extensions}
In this section, we demonstrate how to conduct three macro operations, inversion, shift and broadcast entirely within the subarray, without data leaving the memory bank or requiring dedicated shifting logic in the BLSAs.

\subsection{In-Subarray Inversion (\texttt{RowInvert})}
\label{sec:rowinvert}
\texttt{RowInvert} repurposes migration cells~\cite{lu2015improving} to compute the complement of a row's value. The placement and mechanism are the same as the dual-contact cells (DCCs) proposed by~\cite{seshadri2017ambit}. \paperName{} uses two inversion rows to compute a \texttt{RowInvert}: one is placed at the top of the subarray, the other at the bottom. One access transistor connects to \emph{bitline} and the other to $\overline{\emph{bitline}}$, which is extended down to connect to
one side of the migration cell. \autoref{fig:rowShift} shows the cell structure, including the control lines \emph{dWL} and \emph{nWL}: when \emph{nWL} is activated, the migration cell stores the value from $\overline{\emph{bitline}}$; when \emph{dWL} is activated, it stores the value of \emph{bitline}.  \paperName{}'s \texttt{RowInvert} mechanism uses the same \texttt{AAP} sequence, where the sequence of wordline activations differs based on the direction of the shift, the sequence of wordline activations for \texttt{RowInvert} stays the same. A \texttt{RowInvert} requires a single \texttt{AAP} command with a latency of $\sim$50ns. Another \texttt{AAP} command would be needed to recombine and return the inverted data from the inversion cells.



\subsection{In-Subarray Shifting (\texttt{RowShift})}
\label{sec:rowshift}    

A single migration cell row moves data across the subarray boundary, but the outbound and return transfers traverse the same BLSA, so data returns to its original bitline with zero net displacement. Realizing a full in-row \texttt{RowShift} requires two migration cell rows~\cite{tegge2026low}: the source row is split between the top and bottom migration cell rows, then recombined with a second pair of row clones, achieving a one-position shift.

\begin{figure}
    \centering
    \includegraphics[width=0.85\linewidth]{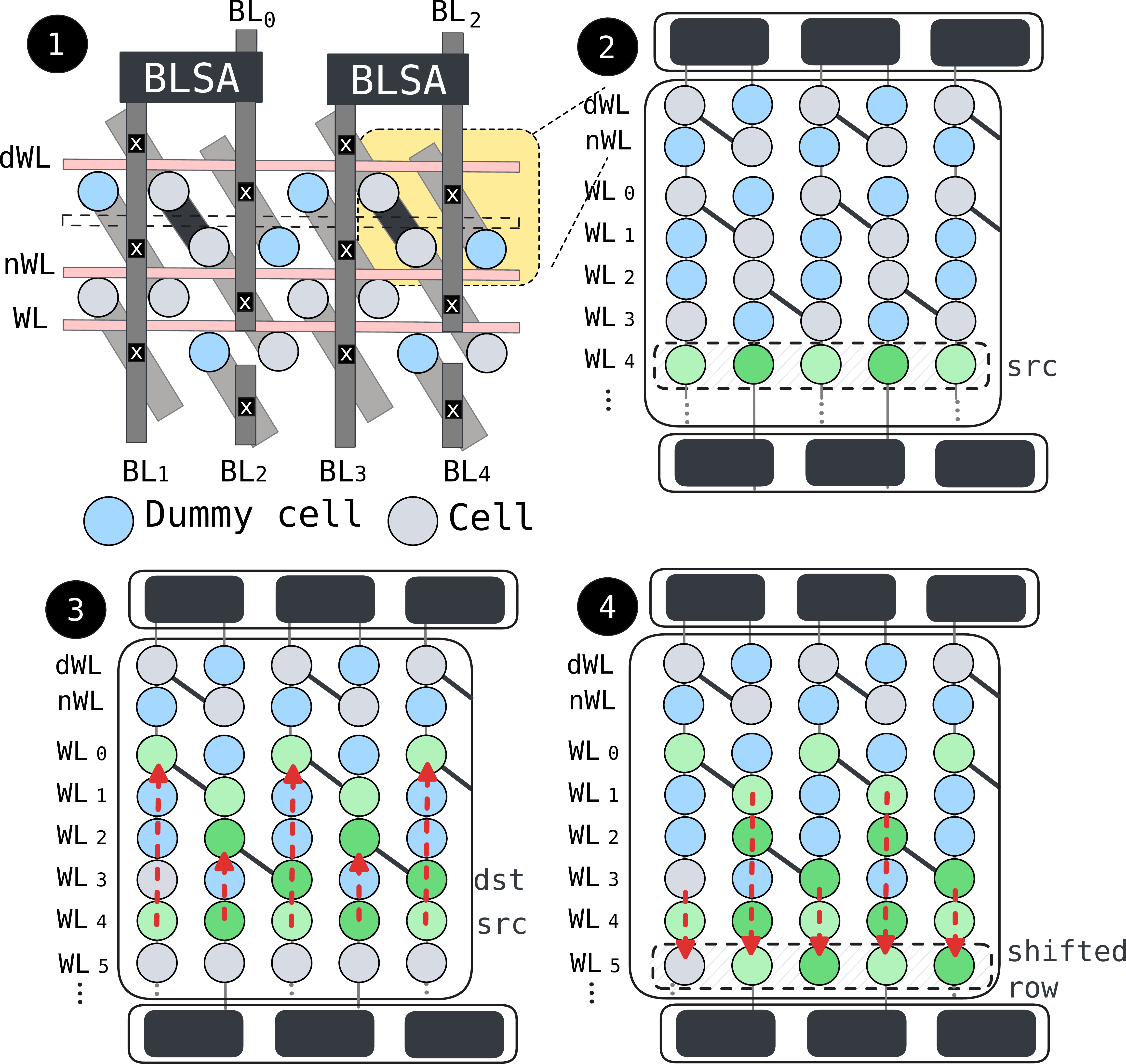}
    \vspace{-1em}
    \caption{Inversion cell structure \ding{182} used in \texttt{RowInvert}; \texttt{RowShift} example using a sequence of RowClone operations \ding{183}--\ding{185}.}
    \label{fig:rowShift}
\end{figure}

\autoref{fig:rowShift} illustrates the steps for a right \texttt{RowShift}: \ding{183}~the initial data on WL4; \ding{184}~\texttt{ACT}(WL4)-\texttt{ACT}(WL0)-\texttt{ACT}(WL2)-\texttt{PRE} splits it across migration rows; \ding{185}~\texttt{ACT}(WL1)-\texttt{ACT}(WL3)-\texttt{ACT}(WL5)-\texttt{PRE} recombines with a one-position offset at WL5. A left shift reverses the activation order. Each \texttt{RowShift} requires 4 \texttt{AAP} commands. The first two can execute in parallel, but the recombine cannot: activating both MC wordlines together would open the dummy cells and dump their charge onto the destination bitlines. The four row clones therefore execute as three sequential steps, with a total latency of $\sim$150\,ns.

\subsection{In-Subarray Broadcast (\texttt{RowBroadcast})}
\label{sec:rowbroadcast}

Beyond shifting and inversion, migration cells enable a third primitive, \texttt{RowBroadcast}, which replicates a single bit across adjacent column positions within a packed element. Where \texttt{RowShift} moves all data in a row by one column, \texttt{RowBroadcast} isolates a bit with an \texttt{AND} against a positional mask and propagates it across a contiguous range of columns, as partial-product generation, conditional selection, and masking require.

\begin{figure}
    \centering
    \vspace{-1em}
    \includegraphics[width=\linewidth]{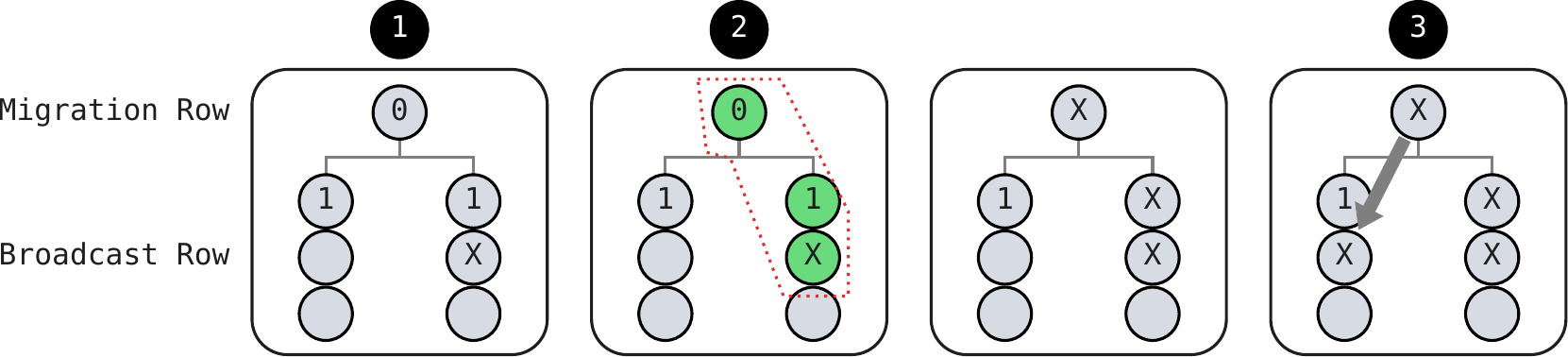}
    \vspace{-2em}
    \caption{Initial state \ding{182} for a \texttt{RowBroadcast}; \ding{183} shows a \texttt{MAJ3} operation using the Migration Cell; \ding{184} \texttt{RowClone} to the Broadcast row.}
    \label{fig:rowbroadcast}
\end{figure}

\texttt{RowBroadcast} exploits the fact that a migration cell can participate in Boolean computation during the transfer. The migration cell rows are pre-loaded with ones by a \texttt{RowClone} from the control-1 row. Each hop then performs a triple-row activation $\texttt{MAJ3}(0, 1, X)$ over the pre-loaded migration cell, the control-0 row, and the source row holding the bit. Since $\texttt{MAJ3}(0, 1, X) = X$ for any $X$, the source row is self-preserving, and the migration cell now holds $X$ and clones it to the neighboring column before being reset to zero for the next hop. \autoref{fig:rowbroadcast} shows these three steps.

Filling all $n$ data columns takes $n - 1$ hops, alternating between the top and bottom migration cell rows. The \texttt{MAJ3} identity preserves already-filled columns, so the bit accumulates in place and no \texttt{OR} merging is required. The first hop costs four operations and each subsequent hop costs three, because the reset ending one hop serves as the pre-load for the next, giving $3n{-}2$ operations per bit. Each of the $n$ multiplier bits needs its own broadcast, so partial-product generation costs $O(n^2)$.

Both DRISA~\cite{li2017drisa} and an alternative doubling strategy instead broadcast via $\lceil \log_2 n \rceil$ shift-then-\texttt{OR} stages at increasing distances, adding $n \lceil \log_2 n \rceil$ \texttt{OR} operations to merge the shifted copies. \texttt{RowBroadcast} eliminates these, and unlike DRISA needs no shift-array hardware beyond the migration cells already present for \texttt{RowShift}.

\subsection{Limitations}
\label{sec:limitations}
\rev{
\paragraph{Shift distance} \texttt{RowShift} moves one column per operation, so a shift of distance $d$ costs $d$ hops. Kogge--Stone keeps logarithmic logic depth but accumulates linear shift cost, so realized latency grows linearly (\autoref{sec:costmodel}). Partial-product masks cost $O(n^2)$ for the same reason, since every multiplier bit must traverse the full element width (\autoref{sec:rowbroadcast}). A \texttt{RowShift} is still far cheaper than a full-adder step, so prefix addition wins at the wider widths we evaluate. At narrow widths the shift term dominates and the compiler selects ripple-carry on cost (\autoref{sec:compiler}).

\paragraph{Cell-array modification} PRADA~\cite{prada} raises two objections to migration cells: (i) skipping a wordline pattern creates a discontinuity in the buried-channel array transistor process, and (ii) cells sharing a migration wordline cannot serve as normal cells. 
We model (ii) directly rather than abstracting it away, which is why \texttt{RowShift}’s recombine step is serialized (\autoref{sec:rowshift}). Objection (i) is a manufacturing question we cannot resolve; we instead quantify the requirement and observe that \paperName{} outperforms a periphery shifter~\cite{li2017drisa}.
Our inversion rows share the placement and mechanism of Ambit's dual-contact cells~\cite{seshadri2017ambit}, so they inherit PRADA's objections to that structure as well. Those are reference bitlines routed through the sense-amplifier region, a break in the gate-bitline pattern, and coupling to the opposite bitlines when \emph{nWL} is enabled.

\paragraph{Designated rows} \paperName{} dedicates two migration and two inversion rows per subarray. These are fixed at design time, cannot serve as storage, and require decoder support, a cost PRADA identifies for any design using designated rows. Provisioning repair rows against faults doubles it. The resulting die area is reported in \autoref{sec:experimental-setup}.

\paragraph{Validation} PRADA observes that such cells' area cost is routinely underestimated and unvalidated in production. We address the first by not relying on a model alone. Our figure is backed by a DRC/LVS-clean transistor-level layout with post-layout extraction and Monte Carlo across process variation (\autoref{sec:layout-validation}). The second applies equally to every DRAM-PUM proposal that modifies the array, including Ambit and DRISA.

}

\section{\paperName{} Bit-Parallel Computation}

Many PUM applications (e.g., HDC, encryption, hashing, database filtering) are dominated by bulk bitwise logic with occasional permutations or rotations, which \texttt{RowShift} makes cheap in both layouts. Arithmetic differs because multiplication requires partial-product generation, carry propagation, and multi-operand reduction across bits of differing significance. BS layouts resolve these dependencies for free via row-index remapping, whereas bit-parallel layouts need explicit horizontal communication for masks and intermediates. Rather than build on sequential carry propagation, \paperName{} composes arithmetic from highly parallel operations and structured horizontal communication, enabling efficient reduction and accumulation while preserving compatibility with bulk bitwise workloads.

\subsection{Element-Wise Kogge--Stone Addition (KSA)}
\label{sec:rapid-addition}

\begin{figure}[t]
    \centering
    \begin{subfigure}{\linewidth}
        \centering
        \includegraphics[width=\linewidth,trim={0cm 1cm 0cm 2cm}, clip]{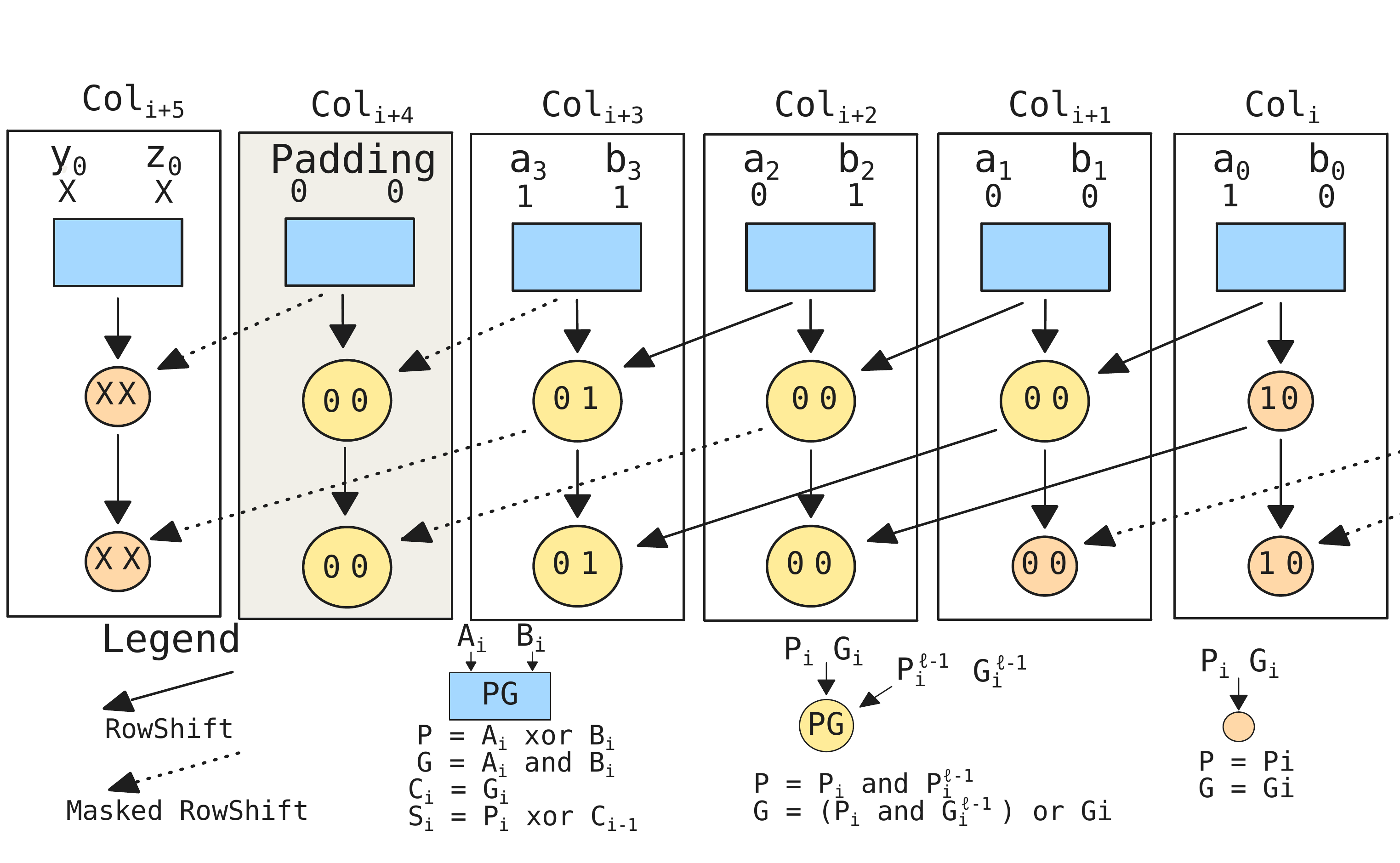}
        \caption{Kogge--Stone addition with operands $A$ = `1001' and $B$ = `1100'.}
        \vspace{0.3cm}
        \label{fig:ksa}
    \end{subfigure}
    \begin{subfigure}{\linewidth}
        \centering
        \includegraphics[width=\linewidth]{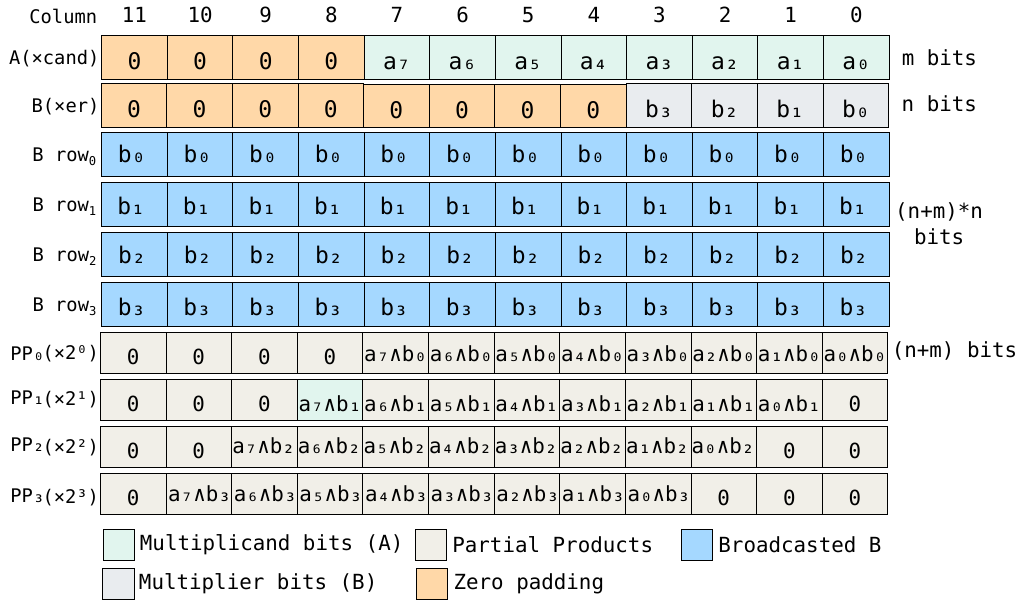}
        \caption{CSA-based parallel multiplication between \texttt{int8} and \texttt{int4} operands.}
        \label{fig:parallel_mul}
    \end{subfigure}
    \caption{Mapping parallel algorithms in horizontal layout.}
    \label{fig:mapping_KSA_mult}
\end{figure}

\autoref{fig:ksa} illustrates the mapping of the KS prefix construction (\autoref{sec:ks-background}) to \texttt{RowShift} and bitwise logic (i.e., \texttt{xor}, \texttt{or} and \texttt{and}).\footnote{\texttt{AP}/\texttt{AAP} synthesis from bitwise primitives is outside this paper's scope and has been addressed by prior works~\cite{hajinazar2021simdram,guo2025pimsynth,Meta-PIM}.}
In the running example in \autoref{fig:ksa}, the inputs $A$ and $B$, both \texttt{int4}, are aligned column-wise and zero-padded to produce an \texttt{int5} output.
Each prefix level's shifted operand $G_{i-2^{\ell-1}}^{(\ell-1)}$ is obtained by shifting the $G$ row leftward by $2^{\ell-1}$ positions via $2^{\ell-1}$ repeated \texttt{RowShift} hops. The \texttt{AND} and \texttt{OR} operations in
the prefix recurrence of~\autoref{sec:ks-background} are realized as $\text{AND}(x,y) = \texttt{MAJ3}(x,y,0)$ and $\text{OR}(x,y) = \texttt{MAJ3}(x,y,1)$.

A subtlety arises from \texttt{RowShift} behavior: unlike a zero-filling barrel shifter, the migration cell has no external data source beyond the subarray boundaries, so a \texttt{RowShift} leftward by $d$ positions fills the rightmost $d$ columns with duplicated boundary values. To ensure correctness, \texttt{RowShift} results are cleaned by AND-masking with a \emph{clear mask} (for $G$, requiring OR-identity 0) or OR-masking with a \emph{set mask} (for $P$, requiring AND-identity 1). The masking is represented in \autoref{fig:ksa} as dashed arrows. These masks are generated once from two seed rows and reused across all arithmetic at that precision. Level-0 signals are extracted using a full adder: $\text{FA}(A,B,0)$ produces both $G = \texttt{MAJ3}(a,b,0)$ and $P = \texttt{MAJ5}(a,b,0,\overline{G},\overline{G})$ in a single step (see \autoref{sec:bg-pim}).


Although KSA holds the logic depth of an individual addition to $\lceil\log_2 n\rceil$ levels, its realized cost is linear in $n$ because each prefix level at distance $2^{\ell-1}$ requires $2^{\ell-1}$  \texttt{RowShift} operations. For multiplication, performing $n$ sequential KS additions would yield $O(n^2)$ total cost. The following section introduces carry-save reduction, which compresses the partial products in $O(n)$ with a single \texttt{RowShift} per stage, followed by one KS addition.

\subsection{Bit-Parallel Element-Wise Multiplication}
\label{sec:rapid-mult}

Binary multiplication of an $m$-bit multiplicand $A$ by an $n$-bit multiplier $B$ decomposes into a sum of $n$ shifted partial products (PP)~\cite{parhami2010computer}. Each $\text{PP}_i$ is formed by conditionally replicating $A$ based on bit $i$ of $B$, shifted left by $i$ positions.
\autoref{fig:parallel_mul} shows the resulting DRAM layout. The masking step requires each multiplier bit to be present in every column of its partial product: bit $b_i$ is isolated by an \texttt{AND} with a positional mask, replicated across the $m+n$ memory columns by \texttt{RowBroadcast} (\autoref{sec:rowbroadcast}), and \texttt{AND}'d with $a$. \texttt{RowShift} then aligns the result to bit position $i$.

The key challenge is efficiently reducing $n$ partial products to a single result. A naive approach, i.e., sequentially reducing partial products using KSA, requires $n-1$ carry-propagating additions at $O(\log n)$ depth each, yielding $O(n \log n)$ total depth. \paperName{} instead exploits carry-save addition (CSA) to eliminate intermediate carry propagation entirely, deferring it to a single final addition.

\subsubsection{Multi-Operand Reduction}
Carry-save addition (CSA) compresses three vector operands ($a$, $b$ and $c$) into two (i.e., a partial sum $S$ and a carry $C$) without propagating carries.
Each bit position is processed by a single full-adder step (see \autoref{sec:bg-pim}), and because no bit position depends on another, all positions are processed simultaneously at single depth.
The carry vector $C$ is subsequently shifted left by one position via \texttt{RowShift} to align it with the next-higher significance for the following addition stage.
This single \texttt{RowShift} is substantially cheaper than the multiple \texttt{RowShift} operations required by the KS prefix levels, making carry-save reduction an efficient building block for partial-product compression.

Applying 3:2 CSA reduction iteratively reduces $n$ partial products to two operands in $n - 2$ full-adder stages. Each stage takes three operands and produces two (sum and shifted carry), reducing the operand count by one. Because each stage has unit depth (no carry propagation), the reduction tree operates in $O(n)$ total depth with each step requiring only a constant number of row-wide PUM operations and a single \texttt{RowShift}. 
\rev{A single KS addition resolves the final two operands. Each 3:2 stage costs one full adder and one \texttt{RowShift}, and the final addition is $O(n)$ through its shift term, giving $O(n)$ overall. This improves on the $O(n^2)$ of sequential KSA or RCA: $n-1$ carry-propagating additions at $2(n{-}1)+1$ shift hops each are replaced by $n-2$ stages that shift by one column each.}

\subsection{Implications for GEMV and GEMM}
\label{sec:gemm}

General matrix-vector multiplication (GEMV) computes $\mathbf{y} = A\mathbf{x}$, where $A$ is a $K \times N$ weight matrix and $\mathbf{x}$ is a $K$-element input vector, producing an $N$-element output vector. GEMV is the dominant computation in single-batch LLM decoding, where each token generation requires a full pass through the model's weight matrices with minimal data reuse~\cite{kwon2023efficient}. This memory-bound characteristic makes GEMV a natural target for PUM~\cite{kubo2025mvdram,de2024count2multiply,he2025papi}.
General matrix-matrix multiplication (GEMM), $C = AB$, generalizes GEMV by stacking $M$ input rows into an $M \times K$ matrix $A$, against the same $K \times N$ weight matrix $B$, to produce an $M \times N$ output $C$. Single-batch prefill or batched inference use $M>1$. Our evaluation targets GEMV directly, computing one output row per layer; the $M>1$ case is realized by executing $M$ independent GEMVs in parallel across banks. Each matrix element is represented using $n$ bits.

In a bit-parallel organization, each $n$-bit element is stored horizontally, requiring no bit-level repacking before computation. As with all GEMM forms, rows of $A$ are aligned with rows of $B^T$. An $n$-bit multiplication needs $n$ partial-product masks, one per multiplier bit; constructing them requires horizontal broadcast (\autoref{sec:rowbroadcast}) at cost 
\rev{$O(n^2)$} per row of $B^T$. Since $B^T$'s rows are reused across all $x$ rows of $A$, masks are built once per row of $B^T$ and reused across its $x$ dot products, each requiring only mask application and reduction of $n$ partial products: 
\rev{$O(n^2 + xn)$}, or amortized per multiplication:
\rev{$O\!\left(n + \tfrac{n^2}{x}\right)$}. In return, horizontal layout enables lower-latency reduction via parallel $3:2$ reduction and KS addition. 

\rev{
\paragraph{Padded layout and host access} \paperName{} changes the \emph{stride} between elements, not the layout of an element. Values sit $2n$ bits apart rather than $n$, each still contiguous and in ascending bit order at a naturally aligned address. An ordinary aligned load or store of element $j$ at $\texttt{base} + j\cdot\texttt{stride}$ therefore works unmodified, with no change to the DDR address mapping. This is the substantive difference from a bit-serial layout, where bit $i$ of every element sits in a different row and one element must be gathered from $n$ rows. The stride is an allocation property, chosen by the compiler in the pass that selects the execution mode. We use $2n$ rather than the minimum $2n-1$ so addressing remains a shift rather than a multiply, at 1.5\% of packing density. Two costs follow. Operand arrays occupy $2\times$ the footprint and a host-side scan touches $2\times$ the cache lines. The density loss is already reflected in all reported numbers through $P$. The footprint and locality costs are not, since they fall outside the PUM execution we model.
}

\section{The \paperName{} Compiler}
\label{sec:compiler}

\paperName{} exposes new in-subarray operations, \texttt{RowInvert}, \texttt{RowBroadcast}, and \texttt{RowShift}, alongside the underlying \texttt{MAJ} logic. Each admits either a horizontal or vertical layout, enabling both bit-parallel and bit-serial arithmetic algorithms whose cost depends on operand layout. Selection cannot be made in isolation: an operation's output layout becomes its neighbor's input layout, and a mismatch between producer and consumer introduces a transpose penalty that can outweigh the algorithmic gain. The compiler must therefore select layout and algorithm jointly, rather than independently for each operation.

\rev{Prior PIM compilation frameworks assume a single, fixed layout~\cite{PRIMO,peng2023chopper,oliveira2024mimdram,ma2024pimsab,khan2024cinm}, and do so because their substrates admit only one: bit-serial PUD offers no row-parallel arithmetic to select between.} Proteus~\cite{de2025proteus} chooses among bit-serial and bit-parallel \emph{algorithms} per operation and bit-precision at runtime, but always over vertically organized data. \rev{\paperName{} is the first DRAM-PUM substrate on which both layouts are natively executable, so the joint layout--algorithm choice the \paperName{} compiler makes across the dependence graph of a vectorized loop body does not arise on prior substrates.}

The \paperName{} compiler is implemented as a pass within the standard \texttt{-O3} LLVM pipeline, and therefore supports both C/C++ via clang and PyTorch via torch-mlir~\cite{torchmlir2022}. The pass delimits the \emph{PIM-vectorizable region}, admitting an instruction only when every operand resides in a PIM buffer and is indexed by a multiple of the vector length $N$, similar to~\cite{PRIMO,oliveira2024mimdram}. Instructions that fail this test fall back to the CPU. \autoref{sec:costmodel} then places algorithm and conversion costs on a common scale, and \autoref{sec:dp} assigns an algorithm and an output layout to every node of the dependency graph in a single dynamic program.


\subsection{Data Layout and RowShift Semantics}

\paperName{} stores an $N$-element integer vector in DRAM in one of two layouts: 
\paragraph{Vertical (V)} Bit $b$ of all $N$ elements occupies one DRAM row. A $w$-bit vector therefore spans $w$ rows. This is the natural layout for BS operations, such as the ripple-carry adder (RCA), that processes one bit-plane per step.
\paragraph{Horizontal (H)} All $w$ bits of element $i$ occupy a single DRAM row. BP adders activate one full row per operand, making this layout optimal for wide-word arithmetic.

\rev{Within a subarray, a prefix adder gains nothing in layout~V. Each level combines bit positions $i$ and $i-2^{\ell-1}$, which are distinct rows, so one level costs $w$ activations and the network $O(w\log w)$, worse than ripple-carry's $O(w)$. Proteus~\cite{de2025proteus} recovers the parallelism by spreading operand bits across subarrays, at the cost of remaining bit-sliced. In layout~H all $w$ bits share a row, so a single row-wide \texttt{MAJ} evaluates an entire prefix level.}

The semantics of \texttt{RowShift} depend on the operand layout. Under layout~V, \texttt{RowShift} moves every data element one position ($\pm 1$) in the array, providing the neighbor access needed for stencil computations. Under layout~H, the same physical operation shifts bits within each element, implementing the logical left/right shift of the data word. The layout choice thus selects between the two semantics.

When \texttt{RowShift} is used to shift positions within an array, a BS algorithm (e.g., RCA) is the natural choice, so its operands and output reside in layout~V. The \paperName{} compiler identifies this case from the array-index access pattern (e.g., a $\pm 1$ offset into an $N$-element column), marks the output buffer with a \emph{fixed-vertical} flag, and propagates that flag to every dependent load in the dependency graph before cost evaluation. A consumer of such output that requests layout~H must pay $T_{\text{trans}}(\text{V}{\to}\text{H}, w)$.



\subsection{Per-Operation Cost Model}
\label{sec:costmodel}

Let $t_{\text{MAJ}}$, $t_{\text{NOT}}$, and $t_{\text{SHIFT}}$ be the primitive row-operation latencies.
The latency of each addition implementation on a vector of $N$ $w$-bit elements is:
\rev{$T_{\text{RCA}}(w) = w \cdot T_{\text{FA}}$, $T_{\text{KSA}}(w) = \left[2(w-1)+1\right] t_{\text{SHIFT}} + \left[5\lceil\log_2 w\rceil + 1\right] t_{\text{MAJ}} + 2\,T_{\text{FA}} + t_{\text{ROWCLONE}}$, and $T_{\text{CSA+KSA}}(w, m) = m\,T_{3:2} + t_{\text{SHIFT}} + t_{\text{MAJ}} + T_{\text{KSA}}(w)$,}
where $m$ is the number of operands in a reduction,
$t_{\text{MAJ}} \equiv t_{\text{MAJ3}}$,
$T_{\text{FA}} = t_{\text{MAJ3}} + t_{\text{NOT}} + t_{\text{MAJ5}}$
is the single-step full adder, and
$T_{3:2} = t_{\text{SHIFT}} + t_{\text{MAJ}} + T_{\text{FA}}
+ 2\,t_{\text{ROWCLONE}}$
is one 3:2 carry-save stage including the carry realignment shift
and its cleanup.

\paragraph{Layout conversion} \paperName{} performs no in-DRAM transposition: converting a
$N{\times}w$ operand between layouts is a host-side operation. The operand is read out
over the memory bus, its bit-matrix is transposed by a CPU routine, and the converted
result is written back. Both terms scale with the operand footprint
$B(N,w) = N w / 8$ bytes:
$T_{\text{xpose}}(N,w) = t_{\text{xpose}} \cdot B(N,w)$, $T_{\text{xfer}}(N,w) = B(N,w) \,/\, \mathit{BW}_{\text{DRAM}}$, and $T_{\text{trans}}(N,w) = T_{\text{xpose}}(N,w) + T_{\text{xfer}}(N,w)$,
where $t_{\text{xpose}}$ is the measured cost of the CPU bit-transposition
routine and $\mathit{BW}_{\text{DRAM}}$ is the host--DRAM bandwidth.
Because the conversion is a host-side bit permutation, its cost is independent of the
direction of travel, so $T_{\text{trans}}(\text{V}{\to}\text{H}) = T_{\text{trans}}(\text{H}{\to}\text{V})$;
we therefore write $T_{\text{trans}}$ without a direction argument. Note that
$T_{\text{xpose}}$ dominates $T_{\text{xfer}}$ by $8\times$ at this bandwidth: layout
conversion is limited by CPU bit-manipulation, not by the memory bus.

$T_{\text{trans}}$ is what makes the \emph{effective} cost of an algorithm depend on its producers: arithmetic algorithms must be compared not in isolation, but including the cost of bringing their inputs into the required layout. As a concrete example, $T_{\text{KSA}}(32)$ is $2.4\times$ cheaper than $T_{\text{RCA}}(32)$ when inputs are already in layout~H. However, if an operand is the output of a bit-serial algorithm and therefore leaves data in layout~V, a KSA consumer must additionally pay a layout conversion. For our $N = 65{,}536$ configuration,
an \texttt{int32} operand occupies $B = 256\,\text{KiB}$, so 
$T_{\text{trans}} \approx 73~\mu s$ on DDR3 (\autoref{tab:ArchParameter}), making the total KSA path $28\times$ \emph{more} expensive than RCA despite KSA's superior arithmetic throughput.

\subsection{Global Optimization via Dynamic Programming}
\label{sec:dp}
Inspired by code generators that use dynamic programming (DP)~\cite{aho1989code}, we formulate layout and algorithm selection as a global cost-minimization problem. 
Let $\mathcal{G} = (\mathcal{V}, \mathcal{E})$ be the dependency tree of the PIM-vectorizable instructions in a loop body. Each node $v \in \mathcal{V}$ represents an $N$-wide vector operation; edges encode def-use dependencies. The algorithm proceeds in two passes:
(1) a bottom-up traversal fills $\mathit{dp}[v][\lambda]$ for every node and every candidate layout $\lambda$;
(2) a top-down traversal propagates the layout selected at the root back through the graph, fixing the concrete instruction at each node.

\textbf{State.}
For each node~$v$ and candidate output layout $\lambda \in \{\text{V},\text{H}\}$, define $\mathit{dp}[v][\lambda]$ as the minimum total latency of the sub-tree rooted at~$v$ when $v$'s result is in layout~$\lambda$.

\textbf{Recurrence.} 
Let $\mathcal{C}(v)$ be the children of~$v$ in $\mathcal{G}$, and let $c(v,\lambda)$ be the cost of the cheapest arithmetic alternative for~$v$ that produces output in layout~$\lambda$. For example, the alternatives for an \texttt{add} node are KSA, CSA, and CPU as fallback\footnote{We assume the CPU fallback requires inputs and produces outputs in layout~H.}. Let $\mu \in \{\text{V},\text{H}\}$ be the layout in which child~$u$ materializes its result.
The DP chooses $\mu$ independently per child to minimize the child sub-tree cost plus any transposition needed to match the parent's required layout~$\lambda$:


\begin{equation}
\mathit{dp}[v][\lambda]
  = c(v,\lambda)
  +\sum_{u \in \mathcal{C}(v)}
    \min_{\mu \in \{\text{V},\text{H}\} }\Bigl(
      \mathit{dp}[u][\mu]
      + T_{\text{trans}}(\mu \to \lambda)
    \Bigr)
  \label{eq:dp}
\end{equation}
where $T_{\text{trans}}(\mu \to \lambda)$ 
abbreviates $T_{\text{trans}}(N,w)$ 
when $\mu \neq \lambda$, and is $0$ otherwise.



\textbf{Output.}
Once the bottom-up fill completes, the root layout $\lambda^* = \arg\min_\lambda \mathit{dp}[\text{root}][\lambda]$ is fixed, and the top-down pass propagates the induced layout choices back through the graph.
The result is a layout-annotated graph in which each node carries a chosen arithmetic implementation and an output layout, and every mismatched edge is annotated with a transposition call. This graph is lowered directly to the PIM instruction stream.

\rev{\textbf{Static selection.} \paperName{} selects layouts at compile time. The element stride is fixed when a buffer is allocated (\autoref{sec:gemm}), and the DP's inputs, operand width and dependence structure, are known statically for the workloads we evaluate. Layout adaptation is not algorithm adaptation. Re-deciding a layout means converting data already held in it, so the gain must cover a transposition, while Proteus chooses an algorithm per operation at no such cost. A workload whose best layout depends on runtime values would need both paths emitted or a conversion at the switch, which we do not evaluate.}

\section{Experimental Setup}
\label{sec:experimental-setup}

To evaluate the impact of \paperName{} we conducted detailed architecture simulations based on layout-level implementations of DRAM circuits, PUM-based simulation, and execution on multiple embedded-class architectures.

\subsection{Layout Evaluation of Migration and Inversion Cells}
\label{sec:layout-validation}
We developed a transistor-level layout in Cadence Virtuoso using the FreePDK \SI{45}{\nano\meter} technology, including BLSAs, migration and inversion cells, conventional DRAM cells, and associated routing. Post-layout parasitic extraction provided the resistance and capacitance values used for SPICE simulation. FreePDK has no trench capacitors, so we substituted metal-insulator-metal capacitors matched to published \SI{45}{\nano\meter} DRAM designs~\cite{tegge2026low} and scaled the bitline and wordline geometries to reported interconnect parasitics~\cite{zhao2006new,cao2011predictive,websitekey}. \rev{FreePDK is a logic process, so the layout evaluates the embedded DRAM path of~\autoref{sec:bg-pim}. The circuit results are calibrated to DRAM process parameters and apply to a commodity 6$F^2$ array, where the angled active area and buried wordline are already in production~\cite{schloesser20086f,Park201520nmDA}.} \autoref{tab:parameters} summarizes the calibrated parameters, with bitline and wordline $R$ and $C$ given per cell. At 512 cells per bitline the total bitline capacitance is ${\approx}143$\,fF ($C_{\mathrm{BL}}/C_{\mathrm{cell}} \approx 4.8$), consistent with prior works~\cite{vogelsang2010understanding,lee2017design}. 
\rev{This ratio is conservative relative to contemporary DRAM. PRADA~\cite{prada} reports $C_{\mathrm{BL}}$ of three to four times $C_{\mathrm{cell}}$ near \SI{20}{\nano\meter} in a vendor's production technology and adopts $4\times$ throughout.} \rev{Because sense margin depends on bitline loading, we sweep $C_{\text{BL}}/C_{\text{cell}}$ rather than relying on a single point. Charge sharing over an $n$-row activation leaves the bitline at $\Delta V = C_{\text{cell}} V_{DD}(k - n/2)/(C_{\text{BL}} + n C_{\text{cell}})$ for $k$ cells holding $V_{DD}$. The worst case $|k - n/2| = 1/2$ is common to \texttt{MAJ3} and \texttt{MAJ5}, but \texttt{MAJ5} carries the larger denominator and is therefore the binding primitive. At our nominal ratio the worst-case \texttt{MAJ5} margin is \SI{61}{\milli\volt}, falling to \SI{46}{\milli\volt} at $8\times$ (equivalent to doubling the bitline to 1024 cells) and \SI{35}{\milli\volt} at $12\times$, remaining above a typical \SI{25}{\milli\volt} sense-amplifier offset throughout. The design is not sensitive to this parameter across the range contemporary DRAM spans.} 
Monte Carlo analysis of the extracted layout (400{,}000 trials, device dimensions varied $\pm\{0, 5, 10, 20\}\%$ of nominal) characterizes the impact of process variation on latency, energy, and reliability, and confirms reliable \texttt{RowInvert} operation and \texttt{MAJ5} resolution within sense-amplifier margins at every variation level. \rev{We additionally model the migration and inversion rows in CACTI using the tool's \SI{22}{\nano\meter} technology file (DDR4-2400, \SI{8}{\giga\bit}), which reports a 1\% die-area overhead.} The resulting timing parameters are incorporated into NVMain-PIM~\cite{poremba2012nvmain,nvmain-pim} for architectural evaluation of PUM performance and energy consumption.

To evaluate the performance of different implementations of bit-parallel and bit-serial computation, we conducted hybrid native execution and PUM simulation using the parameters from \autoref{tab:ArchParameter}. We evaluate two DRAM generations (i.e., DDR3-1333 and DDR4-2400) to span the design points representative of the PUM literature.
DDR3 follows the DDR3-class setup evaluated by Ambit~\cite{seshadri2017ambit}, while DDR4-2400 matches the configuration evaluated by SIMDRAM~\cite{hajinazar2021simdram}. DDR4 is also the mainstream server-memory generation and remains widely deployed in edge-class servers~\cite{HPE-Edgeline,AMD-Versal-UG1273}. 
PUM compute latency changes only modestly across DRAM generations, because it is governed by row-activation timing. In contrast, the transposition overhead that bit-serial architectures incur depends strongly on memory bandwidth. As a result, \paperName{}'s layout compatibility provides a larger advantage on lower-bandwidth platforms (\autoref{sec:eval-e2e}).

We paired an ARM edge-class processor with each memory generation: ARM Cortex-A9 with DDR3 and ARM Cortex-A72 with DDR4. For the bit-serial baselines, data reorganization time, i.e., transposition between bit-parallel and bit-serial layouts, is measured directly on these processor--memory pairs by timing the memory accesses required to redistribute the bits of each matrix element between row-major and column-major organizations. These measurements are performed for each GEMV geometry and reported using the minimum, median, and 95th-percentile latency to capture platform-dependent variability. The ARM platforms are used only for these transposition measurements; the host-side code generated by the \paperName{} compiler is simulated in gem5 on the in-order RISC-V core listed in \autoref{tab:ArchParameter}.


\begin{table}[tbp]
\caption{Technology parameters used in Monte Carlo LTspice simulations.}
\label{tab:parameters}
\centering
\renewcommand{\tabcolsep}{2.5pt}
\footnotesize
\begin{tabular}{lccccccc}
\toprule
 & & & \multicolumn{2}{c}{BL \rev{(per cell)}} & \multicolumn{2}{c}{WL \rev{(per cell)}} & \\
\cmidrule(lr){4-5}\cmidrule(lr){6-7}
Parameter &
$V_{\mathrm{DD}}$ &
Cell Cap &
$R$ &
$C$ &
$R$ &
$C$ &
$t_{\mathrm{rise}}$ \\
\midrule
45\,nm &
1.2\,V &
30\,fF &
200\,m$\Omega$ & 
0.28\,fF &
4\,$\Omega$ &
1.6\,fF &
0.3\,ns \\
\bottomrule
\end{tabular}
\end{table}

\begin{table}[t]
\centering
\caption{Memory organization and architectural parameters.}
\label{tab:ArchParameter}
\footnotesize
\setlength{\tabcolsep}{3pt}
\begin{tabular}{c|p{0.8\columnwidth}}
\toprule
\multirow{9}{*}{DRAM}
& DDR3-1333, 2 ch, 2 ranks/ch, 8 dev/rank, 64-bit bus; 4Gb chip, 8 banks, 1\,kB row, 65536 rows/bank \\
& Timing (ns): $t_{RC}$=49.5, $t_{RAS}$=36, $t_{RP}$=13.5, $t_{FAW}$=30 \\ \cline{2-2}
& DDR4-2400, 2 ch, 2 ranks/ch, 8 dev/rank, 64-bit bus; 8Gb chip, 16 banks, 1\,kB row, 65536 rows/bank \\
& Timing (ns): $t_{RC}$=46.7, $t_{RAS}$=32.5, $t_{RP}$=14.2, $t_{FAW}$=21.7 \\
\midrule
\multirow{3}{*}{\shortstack{host\\CPU}}
& Simulated (gem5): in-order RISC-V @1.45\,GHz, 64\,KB D-cache, 16\,KB I-cache~\cite{sifive2017coreip,lowe2020gem5} \\
& Memory controller, 8\,kB row size, FR-FCFS scheduling \\
& Transposition measured on: ARM Cortex-A9 (DDR3), ARM Cortex-A72 (DDR4) \\
\bottomrule
\end{tabular}
\end{table}

\subsection{Compilation and Simulation Methodology}
\label{sec:scaling}
\paragraph{Compilation pipeline} Workloads are expressed in PyTorch and lowered to LLVM IR through torch-mlir~\cite{torchmlir2022}. Each model is imported into the Torch dialect and lowered to the linalg-on-tensors representation. After bufferization, the linalg operations, i.e., \texttt{linalg.matmul} for GEMV operation and element-wise \texttt{linalg.generic} operations, are converted into explicit loop nests over memrefs, lowered to the LLVM dialect, and emitted as LLVM IR by \texttt{mlir-translate}. The \paperName{} compiler pass (\autoref{sec:compiler}) runs on this IR within the standard \texttt{-O3} pipeline: it delimits the PIM-vectorizable region, jointly selects layouts and algorithms (\autoref{sec:dp}), and lowers each selected operation to its in-DRAM command sequence. Operations outside the region, such as softmax and normalization, remain host code. Because the pass operates on LLVM IR rather than on a framework-specific dialect, the same pass also compiles C/C++ kernels through clang.

\paragraph{Per-operation cost} For each PUM variant and operand width $n \in \{4,8,16,32\}$, the DRAM command sequence of a single $n$-bit multiplication and addition is simulated in NVMain-PIM~\cite{poremba2012nvmain,nvmain-pim} using the timing parameters of \autoref{tab:ArchParameter}, yielding the latency and energy of each operation. Comparing multi-operation traces against composed single-operation results shows less than 0.15\% error in both latency and energy. As in Ambit~\cite{seshadri2017ambit} and SIMDRAM~\cite{hajinazar2021simdram}, workload performance is composed from per-operation costs, which here are obtained from cycle-accurate simulation rather than analytical command counting.

\paragraph{Layer-level composition} A GEMV operator with $N$ output columns and $K$ reduction steps packs $P$ elements per row and distributes its $\lceil N/P \rceil$ tiles across $B$ independent banks~\cite{li2017drisa,oliveira2024mimdram}, giving $T_{\text{layer}} = T_{\text{mask}} + \left\lceil \tfrac{\lceil N/P \rceil}{B} \right\rceil K \left(T_{\text{mult}} + T_{\text{add}}\right)$, where $T_{\text{mask}}$ is the one-time mask derivation cost.
Bit-serial layouts pack $P = W = 65{,}536$ elements per row. 
\rev{Bit-parallel layouts pack $P = \lfloor W/2n \rfloor$, since each element occupies a $2n$-bit field spanned by the shifted partial products during multiply accumulation (\autoref{sec:rapid-mult}), with host access unaffected (\autoref{sec:gemm}), e.g., $P = 1{,}024$ at 32 bits.}
We evaluate single-output-row GEMV ($M = 1$); for $M > 1$, additional output rows occupy idle packed positions at no additional per-row cost (\autoref{sec:amortization}).

\subsection{Layout-Selection Kernels}
\label{sec:dp-bench}
We evaluate fifteen 32-bit integer kernels (\autoref{tab:dp-bench}), compiled from C through clang and the \paperName{} compiler (\autoref{sec:compiler}) on 8{,}192-element vectors: five fused pipelines whose adjacent stages prefer different layouts, and ten single-purpose kernels (stencils, a prefix sum, multiply/add chains). Three properties make the choice non-trivial: neighbour accesses $x_{i\pm d}$ lower to \texttt{RowShift}s whose result the compiler treats as vertical data; bitwise operators exist only vertically; and scalar multiplication only horizontally; element-wise additions and multiplications exist in both forms.

\begin{table}[t]
\centering
\caption{Layout-selection kernels, all on 32-bit integers. $\wedge$ is bitwise AND; $\alpha,\beta,\gamma,c_0,c_1,c_v,w_k,x_k$ are scalars and the other symbols are arrays indexed by $i$. Kernels above the rule are fused pipelines. \texttt{scan} is a four-step Hillis--Steele prefix sum followed by a forward difference.}
\label{tab:dp-bench}
\footnotesize
\setlength{\tabcolsep}{4pt}
\begin{tabular}{@{}l l@{}}
\toprule
Kernel & Expression \\
\midrule
\texttt{jacobi\_w}     & $y_i = c_0\,x_i + c_1\,(x_{i-1}+x_{i+1}) + f_i$ \\
\texttt{blur\_mask}    & $y_i = \big((x_{i-1}+x_i+x_{i+1}) \wedge m_i\big)\,w_i + b_i$ \\
\texttt{qlayer}        & $y_i = \big((a_i u_i + b_i v_i) \wedge m_i\big) + z_i$ \\
\texttt{grad}          & $y_i = \alpha\,(x_{i+1}-x_{i-1}) + \gamma\,x_i$ \\
\texttt{stencil\_axpy} & $y_i = \alpha\,(x_{i-1}+x_i+x_{i+1}) + b_i$ \\
\midrule
\texttt{stencil1d}     & $y_i = x_{i-1}+x_i+x_{i+1}$ \\
\texttt{stencil2d}     & $y_{r,i} = x_{r,i}+x_{r,i\pm1}+x_{r\pm1,i}$ \\
\texttt{scan}          & $x^{(s+1)}_i = x^{(s)}_i + x^{(s)}_{i-2^s}$, $s=0..3$ \\
                       & $d_i = x_{i+1}-x_i$ \\
\texttt{chain}         & $y_i = (a_i b_i + c_i d_i)\,e_i$ \\
\texttt{fmatree}       & $y_i = \sum\nolimits_{k=0}^{3} w_k\, p_{k,i}\, q_{k,i}$ \\[3pt]
\texttt{gemv\_dyn}     & $y_i = \sum\nolimits_{k=0}^{7} x_k\, A_{k,i}$ \\[3pt]
\texttt{bgk}           & $y_i = \alpha\,g_i + \beta\,f_i$ \\
\texttt{eos}           & $P_i = \rho_i T_i,\;\; e_i = c_v\,T_i$ \\
\texttt{euler\_ode}    & $u_i \leftarrow u_i + a_i u_i + b_i$ \\
\texttt{leapfrog}      & $v_i \leftarrow v_i + F_i,\;\; x_i \leftarrow x_i + v_i$ \\
\bottomrule
\end{tabular}
\end{table}

\subsection{Application Workloads}
We evaluate \paperName{} on GEMV operators extracted from 19~machine learning models spanning the MLPerf datacenter and edge inference suites~\cite{reddi2020mlperf}. From each model we extract all linear layers and their dimensions ($N$, $K$) at four operand widths (4, 8, 16, 32~bits), totaling over 20{,}000 layer--width configurations spanning $N$ from 2 to over 40{,}000, $K$ from 4 to over 28{,}000, and six orders of magnitude in per-layer MAC count. For the bit-serial baselines we measure the BP$\leftrightarrow$BS transposition overhead per GEMV geometry on the target platform at minimum, median and 95th-percentile latency, and add it to the bit-serial PUM compute time to obtain end-to-end latency.

\section{Evaluation}

In this section, we evaluate \paperName{} against bit-serial and bit-parallel PUD baselines.
Each configuration is labeled in \autoref{fig:arith} by five orthogonal characteristics:
data layout (BS/BP); logic primitives (AOI, NOR, MAJ3 (MIG), or MAJ3+5 (M35)); inversion
and data-movement mechanism (dual-contact cells, DC; dedicated logic, DL; or
migration-style rows, MR); core adder (ripple-carry, RC, or Kogge--Stone, KS); and
multi-operand reduction (parallel reduction, PR, when present).
\rev{Notable configurations include \textit{BS-AOI-DC-RC} (Ambit-like~\cite{seshadri2017ambit}),
\textit{BS-M35-DC-RC} (SIMDRAM-like~\cite{hajinazar2021simdram} with \texttt{MAJ5}),
\textit{BP-AOI-DL-RC} (\texttt{AND}--\texttt{OR} with ripple-carry),
\textit{BP-NOR-DL-KS-PR} (DRISA-like~\cite{li2017drisa}: \texttt{NOR} logic, a prefix adder,
and carry-save reduction),
\textit{BP-NOR-DL-KS} (DRISA's prefix adder used for shift-and-add multiplication), and
\textit{BP-M35-MR-KS-PR} (\paperName{}).}
We additionally evaluate \textit{BS-M35-MR-KS-PR}, which applies \paperName{}'s arithmetic
optimizations to a bit-serial layout. Comparing it with BP-M35-MR-KS-PR isolates the
impact of bit-parallelism from the algorithmic contributions, while comparisons with the
earlier baselines quantify the benefits of Kogge--Stone addition, parallel reduction, and
majority-logic computation.

\subsection{Core Arithmetic Evaluation}
\label{sec:core-arith}
\newcommand{\change}[1]{%
  \hspace{0.1em}\(\hookrightarrow\)\,#1%
}

\setlength{\tabcolsep}{2.25pt} 

\begin{figure}[tbp]
\centering
\includegraphics[width=\columnwidth,trim={2.4cm 3.3cm 2.5cm 0cm}, clip]{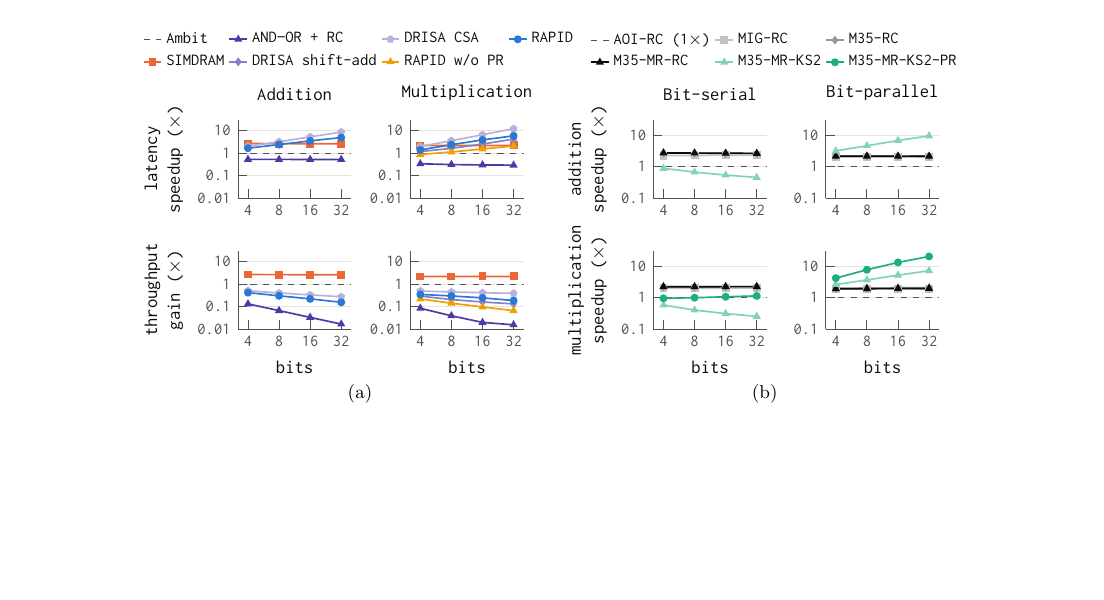}
\vspace{-2em}
\caption{Arithmetic performance across operand widths.
\textbf{(a)}~Improvement over Ambit (BS-AOI-DC-RC).
\textbf{(b)}~Latency speedup of each optimization step over its family's AOI-RC baseline. In the bit-parallel panels, M35-MR-KS2 and M35-MR-KS2-PR are \paperName{} w/o PR and \paperName{} from (a).}
\label{fig:arith}
\end{figure}

\autoref{fig:arith} reports the latency and throughput of each organization for integer addition and multiplication. For bit-serial computation, native majority logic (\texttt{MIG}) provides $\sim2\times$ speedup for addition and 30\%--70\% gains for multiplication over Boolean logic (\texttt{AOI}). Expanding to \texttt{MAJ5} (\texttt{M35}) adds 11\%--14\% for addition and $\sim5\%$ for multiplication. For bit-parallel logic, \texttt{M35} provides a $3\times$ speedup over \texttt{AOI}, making majority logic's benefits more pronounced than in the bit-serial case.

\rev{The configuration \textit{BP-AOI-DL-RC} pairs \texttt{AND}--\texttt{OR} primitives with ripple-carry addition and is not representative of DRISA, which specifies a Kogge--Stone prefix adder and prefers carry-save reduction over full adders~\cite{li2017drisa}. \autoref{fig:arith} therefore reports both corrected configurations. The prefix adder shared by both is 1.22--1.76$\times$ faster than \paperName{}'s, widening with width. DRISA's shift array resolves a prefix level of distance $2^{\ell-1}$ in one operation where \paperName{}'s migration cells require $2^{\ell-1}$ hops (\autoref{sec:limitations}), so the linear shift term of 
$T_{\text{KSA}}$ is paid by \paperName{} alone. At the multiply level \paperName{} is 1.19--1.53$\times$ faster than prefix with shift-and-add but 1.42--2.09$\times$ slower than prefix with carry-save. DRISA is itself bit-parallel and requires no transposition, so the end-to-end comparison of \autoref{sec:eval-e2e} reduces to these per-invocation costs. The distinguishing axis is instead area, quantified in \autoref{sec:isoarea}.}

\subsection{Partial-Product Amortization in BP GEMV}
\label{sec:amortization}

In \paperName{}'s bit-parallel multiply, each reduction step generates $n$ partial-product masks by extracting each bit of the multiplier~$B$ via \texttt{RowBroadcast} and AND-ing with the multiplicand~$A$. This dominates the multiply, growing from 34.4\% of its cost at 4-bit to 69.5\% at 32-bit with the $O(n^2)$ broadcast.

Because every PUM primitive operates on the \emph{entire row simultaneously}, that cost is paid once per reduction step rather than once per element. In the GEMV $\mathbf{y}_{1 \times N} = \mathbf{x}_{1 \times K} \cdot W_{K \times N}$, the $N$ outputs occupy $P$ packed positions of one subarray row, \rev{where $P = \lfloor W_{\text{row}} / 2n \rfloor$ follows from the $2n$-bit element stride (\autoref{sec:gemm}) (e.g., $P = 65{,}536 / 64 = 1{,}024$ at 32-bit).} The scalar $x[k]$ holds the same value at every position, so a single row-wide \texttt{RowBroadcast} fills all $P$ elements at once; the subsequent \texttt{AND} against the weight row $W[k,:]$ then yields a \emph{different} partial product at each position, since each holds a different weight. One broadcast therefore produces $P$ partial products, where without row-wide execution each position would repeat it and multiply the dominant cost by $P$. When $N < P$ only $\min(N,P)$ positions are active; when additional output rows occupy idle positions ($M > 1$) the sharing extends across both dimensions.

Comparing the realized per-step cost $C_{\text{amort}} = C_{\text{PP}} + C_{\text{accum}}$ against a non-amortized model in which each active position builds its own masks, $C_{\text{non-amort}} = \min(N,P) \times C_{\text{PP}} + C_{\text{accum}}$, gives an amortization factor $\alpha = C_{\text{non-amort}} / C_{\text{amort}}$ of $132\times$ on the smallest-$N$ benchmark, rising to over $670\times$ when all \rev{$P = 1{,}024$} positions are utilized (\autoref{tab:amortization}).

This is not an optimization applied to the baseline but an inherent property of row-wide execution, and it is what makes bit-parallel GEMV competitive with bit-serial: for BERT-Large at 32-bit on DDR4, \paperName{} completes all GEMV operators with $6.3\times$ lower end-to-end latency than SIMDRAM including transposition, widening to $97\times$ on the lower-bandwidth DDR3 platform.

\begin{table}[t]
\centering
\caption{Partial-product amortization factor $\alpha$ at 32-bit precision
(\rev{$P=1{,}024$} packed positions). Active positions are $\min(N,P)$;
benchmarks with $N\geq P$ saturate the row and achieve maximum sharing.}
\vspace{-1em}
\label{tab:amortization}
\scriptsize
\setlength{\tabcolsep}{3pt}
\begin{tabular}{@{}lrrr@{\hspace{8pt}} lrrr@{}}
\toprule
Benchmark & $L$ & Max $N$ & $\alpha$ & Benchmark & $L$ & Max $N$ & $\alpha$ \\
\midrule
UNet3D        &    19 &   320  & 132$\times$ & Qwen3-VL 4B    & 359 & 152k  & 673$\times$ \\
RGAT          &   133 &   256  & 156$\times$ & LLaMA-3.1 8B   & 226 & 128k  & 674$\times$ \\
RetinaNet     &    71 &  2.0k  & 236$\times$ & WAN2.2 5B      & 307 & 18.4k & 674$\times$ \\
PointPainting &    62 &  2.0k  & 278$\times$ & Whisper-L v3   & 517 & 51.9k & 674$\times$ \\
ResNet-50     &    54 &  2.0k  & 289$\times$ & LLaMA-2 70B    & 562 & 32.0k & 674$\times$ \\
SDXL          & 1,134 & 49.4k  & 639$\times$ & LLaMA-3.3 70B  & 562 & 128k  & 674$\times$ \\
GPT-OSS 20B   &    98 &  201k  & 648$\times$ & DLRMv3         & 152 & 40.0M & 674$\times$ \\
LLaMA-4 Scout &   386 &  202k  & 659$\times$ & DLRMv2         &  34 & 40.0M & 674$\times$ \\
BERT-Large    &   148 & 30.5k  & 665$\times$ & GPT-J 6B       & 170 & 50.4k & 674$\times$ \\
Mixtral 8$\times$7B & 130 & 32.0k & 672$\times$ & & & & \\
\bottomrule
\end{tabular}
\end{table}

\subsection{End-to-End GEMV Comparison}
\label{sec:eval-e2e}

We compare \paperName{} (BP-M35-MR-KS-PR) against SIMDRAM (BS-M35-DC-RC) and Ambit (BS-AOI-DC-RC) across all benchmarks at 32-bit precision on DDR3, DDR4 platforms. \paperName{} requires no data layout transformation; the bit-serial baselines require bit-parallel to bit-serial transposition before PUM execution and the reverse transposition afterward to return results in CPU-compatible format.

For each benchmark, the \paperName{} compiler analyzes the layer dependence graph and selects bit-parallel execution with row-major data layout for all GEMV operators, confirming that no bit-serial fallback or intermediate transposition is needed for the workloads evaluated.

\rev{
\paragraph{Why conversion is required at every boundary} The costs below assume a BP$\leftrightarrow$BS roundtrip at each GEMV operator. These models separate linear layers by host-resident operations such as softmax, normalization, activations and residual adds (\autoref{sec:scaling}), which read and write row-major. A bit-serial result must return to bit-parallel form for the host and be converted back for the next layer, whereas \paperName{} never leaves the host's layout. Three qualifications bound this. Transposition amortizes with reuse, and we evaluate single-batch decoding ($M{=}1$) where none exists~\cite{kwon2023efficient}. Sweeping $M$, the median benchmark retains 1.6--2.4$\times$ at $M{=}4$ and 1.2--1.7$\times$ at $M{=}8$, crossing over near $M{=}13$--$20$. A model PUM-vectorizable end to end would convert only at the region boundary. None in our suite is, and the advantage would then scale with the boundaries remaining. Finally, our baselines represent the no-overlap case, since we model transposition as a serial host-side step and omit SIMDRAM's cache-side transposition unit and subarray overlap~\cite{hajinazar2021simdram}.
}

\subsubsection{PUM-Only Compute}

For PUM compute alone, \paperName{}'s $O(\log n)$ \rev{logic-}depth Kogge--Stone addition yields 2.2--2.3$\times$ lower latency than SIMDRAM's ripple-carry on benchmarks where $N \leq P$ (single tile), such as BERT-Large, ResNet-50, and PointPainting. For large-$N$ benchmarks (DLRMv2, LLaMA-2 70B), \paperName{} requires multiple tiles ($\lceil N/1{,}024 \rceil$ at 32-bit) while SIMDRAM fits all elements in one row, giving SIMDRAM up to 3.3$\times$ lower PUM-only latency. However, this advantage is reversed when transposition costs are included.

\subsubsection{End-to-End Including Transposition}

Bit-serial PUM requires a BP$\leftrightarrow$BS transposition roundtrip for each GEMV operator, measured per geometry on each DRAM platform.

\paragraph{DDR3} Transposition dominates all other costs. \paperName{} achieves 18.6--96.9$\times$ speedups over SIMDRAM across all benchmarks.

\paragraph{DDR4} Higher bandwidth narrows the gap. \paperName{} achieves 4.1--6.6$\times$ speedups. For LLaMA-2 70B, \paperName{}'s PUM compute is 3.2$\times$ slower than SIMDRAM's, but the transposition roundtrip results in \paperName{} being 6.0$\times$ faster end-to-end.

\begin{table}[t]
\centering
\caption{Transposition overhead as a fraction of PUM compute and its impact on the \paperName{} vs.\ SIMDRAM end-to-end comparison at 32-bit precision across DRAM generations.}
\vspace{-1em}
\label{tab:treorg_trend}
\footnotesize
\begin{tabular}{l|r|r|l}
\hline\hline
Platform & $T_{trans}$ / PUM & \paperName{} speedup & Winner \\
\hline
DDR3 & 25{,}945\% & 94.0$\times$ & \paperName{} \\
DDR4 & 1{,}485\% & 5.9$\times$ & \paperName{} \\
\hline\hline
\end{tabular}
\end{table}


Transposition overhead is therefore the decisive factor in the \paperName{}--SIMDRAM comparison, as \autoref{tab:treorg_trend} summarizes. At lower precisions KS's per-invocation advantage diminishes and SIMDRAM's parallelism gap grows, but transposition still preserves \paperName{}'s end-to-end advantage on both generations.

\subsubsection{Throughput Analysis}


\paragraph{PUM-only throughput} In isolation, SIMDRAM achieves 2.7--2.8$\times$ higher throughput than \paperName{} across platforms, reflecting 
\rev{the 64$\times$ parallelism gap (65{,}536 vs.\ 1{,}024 elements per row at 32-bit)} that offsets \paperName{}'s lower per-invocation latency.

\paragraph{End-to-end throughput} Transposition inverts this comparison on DDR3/DDR4: \paperName{} achieves 5.9$\times$ higher end-to-end throughput on DDR4 and 94$\times$ on DDR3, consistent across benchmarks.

\begin{figure}[t]
\centering
\includegraphics[width=\columnwidth]{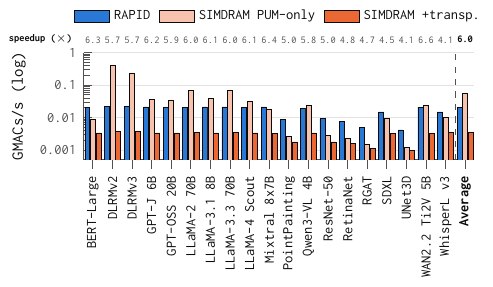}
\vspace{-2.3em}
\caption{End-to-end GEMV throughput at 32-bit precision on DDR4, sorted by \paperName{}'s speedup over SIMDRAM including transposition (right column). Bit-serial throughput includes the measured BP$\leftrightarrow$BS transposition (minimum roundtrip); \paperName{} requires no transposition.}
\label{fig:throughput}
\end{figure}

\paragraph{Implications for system integration} Real ML inference pipelines interleave PUM-accelerated GEMV operators with non-PUM operations such as activations, normalization and softmax on the host, incurring a transposition roundtrip at each of tens to hundreds of transitions per inference pass. \paperName{} eliminates this by preserving native bit-parallel layouts throughout PUM execution. 

\rev{
\paragraph{Iso-area comparison} \label{sec:isoarea} \paperName{}'s extensions cost under one percent of die area (\autoref{sec:experimental-setup}), so a baseline reclaiming that area as additional PUM-capable subarrays barely shifts the comparison against the bit-serial baselines. Our inversion rows are moreover the dual-contact cells Ambit and SIMDRAM already require (\autoref{sec:rowinvert}), so the only rows we add are the two migration rows. The comparison matters far more against DRISA, whose 1T1C variants hold half the capacity of the 8\,Gb reference part at comparable or larger die area, computed from its own reported figures~\cite{li2017drisa}. At iso-area DRISA therefore fits between a third and half as much array as an unmodified die, where \paperName{} fits nearly all of it. Combined with the per-invocation costs of \autoref{sec:core-arith}, this gives \paperName{} 2.4--3.1$\times$ higher throughput per unit area against prefix-adder DRISA with shift-and-add, narrowing to 1.0--1.6$\times$ against carry-save and effectively a tie at \SI{32}{bit}.
}

\subsection{Layout Selection vs. Fixed-Layout Policies}
\label{sec:dp-eval}
The GEMV layers of \autoref{sec:eval-e2e} all resolve to bit-parallel execution, so they never exercise the compiler's choice between layouts. We therefore use fifteen 32-bit integer kernels (\autoref{tab:dp-bench})  to compare the compiler's DP (\autoref{sec:dp}) with two fixed policies: \emph{vertical-only} and \emph{horizontal-only}, which force the vertical or the horizontal implementation wherever an operator offers both. The exception is when an operator offer only one implementation, such as scalar multiplication that exist only horizontally. A fixed plan therefore cannot avoid every conversion. Because the DP minimizes this cost, it is never worse than a fixed policy by construction.

\begin{figure}[t]
\centering
\includegraphics[width=\columnwidth]{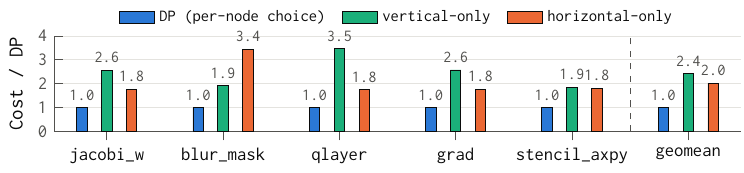}
\vspace{-2em}
\caption{Cost of the fixed layout policies relative to the DP (lower is better) on the five fused pipelines.}
\label{fig:dp-fused}
\end{figure}

\paragraph{Fused pipelines} \autoref{fig:dp-fused} shows that on every pipeline the DP is cheapest and each fixed policy costs between $1.75\times$ and $3.47\times$ more (geometric mean $2.4\times$ for vertical-only and $2.0\times$ for horizontal-only). Neither is best throughout: horizontal-only wins on four pipelines and vertical-only on \texttt{blur\_mask}, where the shifts and the AND are vertical-only and horizontal-only must transpose around them four times. The DP instead keeps the stencil and the bitwise stage vertical and runs the scalar and array multiplies horizontally, transposing V$\to$H once. On \texttt{blur\_mask}, vertical-only transposes as often as the DP yet costs $1.92\times$ more because it relies on serial multiplication.

\begin{figure}[t]
\centering
\includegraphics[width=\columnwidth]{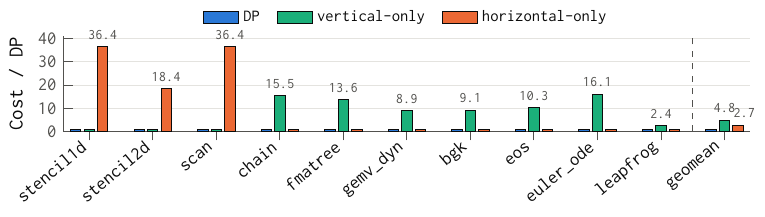}
\vspace{-2em}
\caption{Cost of the fixed layout policies relative to the DP on the single-purpose kernels.}
\label{fig:dp-single}
\end{figure}

\paragraph{Single-purpose kernels} On the ten single kernels (\autoref{fig:dp-single}) the DP equals the better fixed policy, because there a fixed policy is itself the oracle: horizontal-only is right for seven kernels and vertical-only for the three that use shifts. The penalty for choosing wrong is large and asymmetric. Horizontal-only costs $18\times$ to $36\times$ more on the shift kernels, since every shifted operand is vertical data and must be transposed back, and vertical-only costs $2.4\times$ to $16\times$ more on the multiply and add kernels, from serial multiplication and the transpositions around scalar multiplies. The geometric-mean penalties are $4.8\times$ and $2.7\times$. A greedy policy that picks each operator's cheapest implementation without regard to transpositions coincides with horizontal-only on fourteen of the fifteen kernels and inherits its penalty on the shift kernels.

\section{Related Work}
\label{sec:related}


Prior works provide individual
pieces (migration cell, BP algorithms, a way to hide
transposition, or a per-operation algorithm choice), but \emph{none} executes
BP arithmetic on the host's row-major layout without periphery
logic, nor selects layout and algorithm jointly across a program. \paperName{} touches four lines of prior work: 

\paragraph{Horizontal data movement in DRAM}
Migration cells were introduced to move rows between subarray regions of
different latency~\cite{lu2015improving}, not for computation.
DRISA~\cite{li2017drisa} places shifters next to the sense amplifiers, resolving
a power-of-two shift in one operation instead of \paperName{}'s $2^{\ell-1}$ hops
(Sec.~\ref{sec:limitations}), at the cost of logic devices occupying
24--51\% of die area. Other works exploit horizontal movement in other memories~\cite{akin2015data,CORUSCANT,Sieve}.

\paragraph{Array modifications and their cost.} \paperName{}’s inversion rows share the mechanism of Ambit’s DCC ~\cite{seshadri2017ambit,hajinazar2021simdram,oliveira2024mimdram,de2025proteus,oliveira2026pipedram_arxiv,de2024count2multiply}, and migration cells share their structural concerns. PRADA~\cite{prada} notes that such cells break the regular wordline pattern, turn co-located cells into dummy cells, and, for DCCs, carry an underestimated and production-unvalidated area cost. We model the dummy-cell constraint and the designated rows explicitly (Sec.~\ref{sec:limitations}) and validate circuit behavior with a post-layout extracted design. Manufacturing feasibility of the skipped-wordline pattern, as for any array modification, can only be assessed by process engineering.

\paragraph{Bit-parallel arithmetic in memory.}
AritPIM~\cite{leitersdorf2023aritpim} develops BS and BP algorithms for memristive crossbars under NOR/NOT logic, with gate-step costs on a single modeled row that are not commensurable with DRAM row activations. Proteus~\cite{de2025proteus} executes BP algorithms on vertical data by distributing operand bits across subarrays, so its data remain bit-sliced and require transposition at the host. 

\paragraph{Handling the PUM--host layout mismatch.}
SIMDRAM~\cite{hajinazar2021simdram} hides transposition with an LLC-side transposition
unit and subarray overlap; PipeDRAM~\cite{oliveira2026pipedram_arxiv}
permutes bits per memory request at the controller. Both keep BS
compute and optimize the layout boundary. \paperName{} removes the boundary, leaving data at
rest as the host wrote them. The approaches are complementary, as high data
reuse amortizes transposition and favors BS execution
(Sec.~\ref{sec:amortization}), whereas our results model a serial host-side
transposition and thus the low-reuse regime, such as single-batch decoding.

\paragraph{Compilers for PUM}
Prior works (e.g., \cite{peng2023chopper,oliveira2024mimdram,guo2025pimsynth,Meta-PIM,ma2024pimsab,PRIMO} assume a single fixed data layout. Proteus~\cite{de2025proteus} selects an algorithm per operation at runtime but always over vertical
data, so a layout conversion never enters its decision. In \paperName{}'s compiler both layouts are natively executable on the same substrate, the choice of one operation's algorithm fixes its neighbors' input layouts, and the compiler must choose layout and algorithm jointly.

\section{Conclusions}
\paperName{} performs integer arithmetic on bit-parallel, CPU-compatible layouts using migration cells for horizontal movement and inversion cells for in-array \texttt{NOT}. SIMDRAM retains 2.7--2.8$\times$ higher PUM-only throughput, but eliminating transposition at every PUM--host boundary gives \paperName{} 5.9$\times$ higher end-to-end throughput on DDR4 and 94$\times$ on DDR3. Joint layout--algorithm selection lowers latency a further 1.8$\times$.

\bibliographystyle{IEEEtran}
\bibliography{references}

\end{document}